\documentclass[journal]{IEEEtran}
\usepackage{hyperref}
\hypersetup{
     colorlinks   = true,
     citecolor    = blue,
     linkcolor = blue
}

\usepackage{amsfonts}
\usepackage{amsmath, amssymb, bm, amsthm}
\usepackage{braket}

\usepackage{tikz}
\usetikzlibrary{quantikz2}

\usepackage{graphicx}
\usepackage{subcaption}

\usepackage{siunitx}

\newcommand{\Problem}{CDP} %

\newcommand{\vctr}[1]{\boldsymbol{#1}} %
\newcommand{\mtrx}[1]{\mathrm{#1}}
\newcommand{\imagi}{\mathrm{j}}
\newcommand{\Reals}{\mathbb{R}}
\newcommand{\Complex}{\mathbb{C}}

\newcommand{\spatialfreq}[2]{\ensuremath{k_{#1,#2}}}
\newcommand{\steeringvector}{\vctr{a}}
\newcommand{\steeringvectorS}{\steeringvector(s)}
\newcommand{\bigO}[1]{\mathcal{O}{(#1)}}
\newcommand{\covarianceM}{\mtrx{R}}
\newcommand{\covarianceMscaled}{\widetilde{\covarianceM}}
\newcommand{\trace}[1]{\operatorname{tr}(#1)}
\newcommand{\conditionnumber}{\kappa}
\newcommand{\conditionthreshold}{\bar{\kappa}}
\newcommand{\Tr}{\operatorname{Tr}}

\newcommand{\polyP}{\widetilde{P}}

\begin{document}
\title{A Quantum Capon Beamformer for TomoSAR:\\ Super-Resolution Retrieval of Forest Vertical Structure via Quantum Singular Value Transformation}

\author{N.~Petropoulos,
        A.~Semenov,
        N.~Murphy,
        A.~Miroszewski,
        G.~Meoni,
        E.~Blokhina,
        
\thanks{N. Petropoulos, A. Semenov, N. Murphy, and E. Blokhina are affiliated with Equal1 Labs, Dublin, Ireland (e-mail: andrii.semenov@equal1.com; niall.murphy@equal1.com; elena.blokhina@equal1.com).}%
\thanks{G. Meoni and A. Miroszewski are affiliated with the $\Phi$-lab, European Space Research Institute (ESRIN), European Space Agency (ESA), Frascati, Italy.}
\thanks{G. Meoni is also affiliated with the Advanced Concepts and Studies Office, European Space Research Institute (ESRIN), European Space Agency (ESA), Frascati, Italy.}}

\maketitle

\begin{abstract}
Synthetic aperture radar tomography (TomoSAR) reconstructs the vertical distribution of backscattered power within each resolution cell, enabling the separation of ground and canopy returns that is central to forest height and biomass retrieval. The Capon, or minimum-variance distortionless-response, beamformer achieves the elevation super-resolution at the cost of inverting a sample covariance matrix at every pixel. We recast this approach as a complete quantum algorithm, the \emph{quantum Capon beamformer}: the covariance matrix is block encoded, its inverse is realized through a Quantum Singular Value Transformation (QSVT) algorithm, and the Capon power is read out directly from the resulting circuit by a Hadamard test, so that the full pipeline from covariance to power spectrum executes as a single quantum circuit per pixel and height. Validated on ESA AfriSAR P-band data over the Mondah forest against classical Capon and Fourier baselines with airborne LVIS LiDAR as ground truth, the presented algorithm reproduces the classical Capon tomogram with a difference below $0.02$~m between the aggregate peak-height RMSE values against LiDAR and above $0.99$ in structural similarity across the operationally relevant range.
We further identify the QSVT condition number bound as a spectrally selective regularizer parameter. By tuning it, we trace a continuous family of reconstructions between the adaptive Capon and non-adaptive Fourier responses, acting on the spectrum selectively rather than solely through uniform shift of diagonal loading. 
With a logarithmic qubit footprint this method becomes relevant for larger covariance dimensions of related adaptive-filtering and super-resolution problems.  
\end{abstract}

\begin{IEEEkeywords}
SAR tomography (TomoSAR), forest height, image processing, super-resolution, Capon beamforming, quantum signal processing (QSP), quantum singular value transformation (QSVT), minimum variance distortionless responce (MVDR), covariance matrix inversion, Earth observation (EO)
\end{IEEEkeywords}

\IEEEpeerreviewmaketitle

\section{Introduction}

\IEEEPARstart{S}{ynthetic} aperture radar tomography (TomoSAR) extends conventional two-dimensional SAR imaging to the three-dimensional scattering distribution within a resolution cell~\cite{Reigber2000, Tebaldini2012, Fornaro2003}. By processing a stack of $N$ acquisitions taken from different look angles, TomoSAR resolves backscattered power along the elevation direction, perpendicular to the line of sight. This capability is of particular value over forested terrain, where returns from the ground surface and the vegetation canopy are superimposed within a single resolution cell and must be separated to retrieve structural parameters such as canopy height and ultimately, above-ground biomass and carbon stock~\cite{LeToan2011, Quegan2019-ec}. %
At each resolution cell, elevation reconstruction reduces to a spectral estimation problem, %
where the dominant peaks of the estimated power spectrum locate the scattering layers. 
Conventional Fourier beamforming provides a non-adaptive baseline, whose elevation resolution is limited by the available aperture. Adaptive super-resolution estimators overcome this limitation by exploiting the second-order statistics of the data; the most well-known among them is the Capon, or Minimum Variance Distortionless Response (MVDR), beamformer~\cite{Capon1969, LombardiniReigber2003}, which minimizes the contribution of all elevations except the one of interest, while preserving the response in the look direction. %

The Capon estimate is defined by the inverse of the sample covariance matrix $\mtrx{R}\in\mathbb{C}^{N\times N}$, which must be formed and inverted independently for every resolution cell. Given that $\mtrx{R}$ is frequently ill-conditioned due to the closely spaced baselines, this inversion is conventionally stabilized by \emph{diagonal loading}~\cite{Carlson1988}. This is a uniform shift of all eigenvalues that trades elevation resolution for numerical stability across the entire spectrum. This single, spectrally uniform lever is usually the only regularization mechanism employed.

SAR processing involves a substantial amount of linear algebra and signal processing, which makes it a potential good fit for quantum computing. 
There have been some efforts to apply quantum computing to SAR processing directly, although almost entirely at the \emph{image-formation} stage to date. 
Previous work includes quantum formulations of the range-Doppler~\cite{Giovagnoli2025, Salahat2025} and range-migration ($\omega$-$k$)~\cite{Waller2023} focusing algorithms which modify the matched-filtering and Fourier-based steps that turn raw echoes into a focused image, by mapping the fast Fourier transforms at their core onto the quantum Fourier transform. 

The tomographic problem addressed in this work differs in both stage and computational structure, since it operates on a stack of \emph{already-focused} complex images, and its core is the inversion of a per-pixel covariance matrix for adaptive spectral estimation.

Outside image formation, Huber et al.~\cite{Huber2022} considered SAR system design, solving the MVDR beamforming problem for a five-element receive array on a quantum annealer. There, the noise covariance is taken as the identity, so the MVDR weights reduce to those of the conventional beamformer. No covariance inversion is needed and the solution does not depend on the data. 
 
The inversion-dominated adaptive beamforming stage of the SAR chain has to our knowledge not previously been addressed with a quantum algorithm.

In this work, we implement a complete quantum realization of the Capon beamformer (which we call the \emph{quantum Capon beamformer}), within a TomoSAR processing chain, using the Quantum Singular Value Transformation (QSVT)~\cite{Gilyen2019, Martyn2021} framework and validate it on real data against the classical Capon and Fourier beamformers. 
QSVT is a unifying framework that applies a chosen polynomial to the singular values of an embedded (\emph{block-encoded}) matrix into a larger unitary. Matrix inversion is a special case in which the polynomial approximates the function $f(x)=x^{-1}$ over the spectral interval $[-1, -1/\conditionthreshold] \cup [1/\conditionthreshold,1]$, where $\conditionthreshold$ is a chosen bound on the condition number $\conditionnumber$ of $\mtrx{R}$.  
The bound $\conditionthreshold$ thus enters QSVT in a dual role; it bounds the spectral domain on which the inverse polynomial is approximated sufficiently, and it also sets the degree of the approximating polynomial and the size of the quantum circuit.

\begin{figure}
    \centering
    \includegraphics[width=\linewidth]{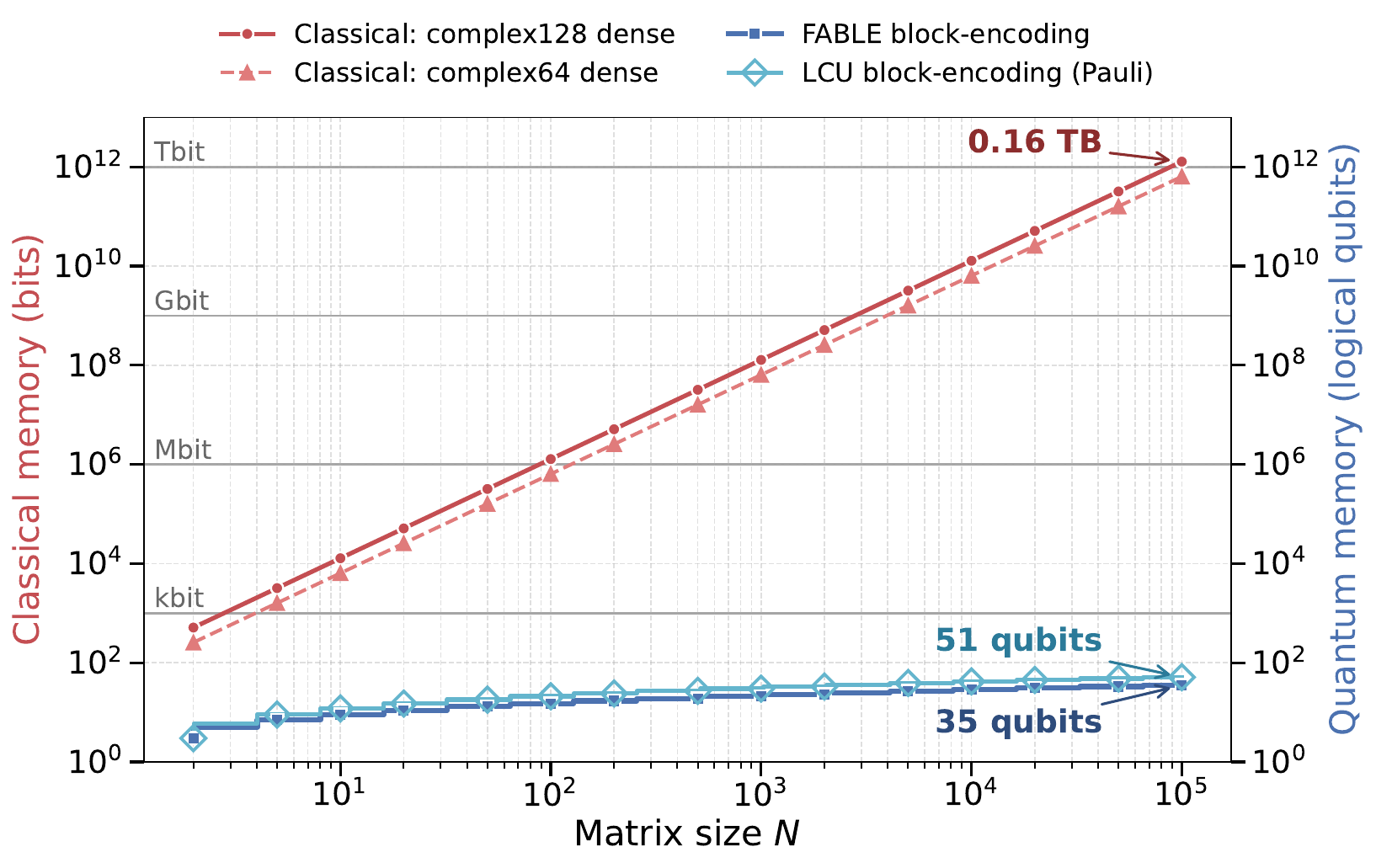}
    \caption{Classical and quantum (QSVT) memory requirements as a function of the covariance matrix dimension $N$ for the TomoSAR problem.
    Classical memory usage (in bits) for the dense covariance matrix are shown for 128 and 64 complex number formats (left axis, red lines).
    The usage of logical qubits for the quantum Capon circuit (right axis, blue lines) are shown for 
    two types of matrix quantum block-encoding techniques are given: FABLE~\cite{campsfable2022} and LCU~\cite{Childs2012}.
    }
    \label{fig:memory}
\end{figure}

A motivating case for QSVT based algorithms for linear algebra is that they scale well with large matrices.
That is, the number of qubits required to implement the quantum Capon beamformer scales logarithmically, $\bigO{\log N}$, in the matrix dimension $N$, as opposed to the  $\bigO{N^2}$ numbers that a naive classical algorithm uses to store a dense matrix (Fig.~\ref{fig:memory}). 
Note that the qubit counts in Fig.~\ref{fig:memory} are \emph{logical} qubits. 
QSVT requires a fault-tolerant quantum implementation where many
physical qubits encode a single \emph{logical} qubit to suppress noise. 
(The number of physical to logical qubits can range from 1 to 3 orders of magnitude depending on the error correction scheme used~\cite{Fowler2012,Bravyi2024}).

Usually $\conditionthreshold$ is fixed once for the sweep while $\conditionnumber$ varies per pixel.
Using ESA AfriSAR P-band acquisitions over the Mondah tropical forest site in Gabon, with co-located LVIS LiDAR canopy height (CHM) and digital terrain (DTM) models~\cite{Blair1999} as ground truth, we show that for a $\conditionthreshold \ge 148$ the QSVT inverter reproduces the classical Capon tomogram to high fidelity across a two-dimensional sweep over track count and condition number bound, measured through peak-height error, center-of-gravity (CoG) bias, and structural-similarity (SSIM) and peak-signal-to-noise (PSNR) metrics between the quantum and classical reconstructions. Furthermore, we formulate the Capon power estimate as a quantity that is directly accessible on a quantum processor. Instead of reconstructing $\mtrx{R}^{-1}$ explicitly, the scalar $\steeringvector^\dagger \mtrx{R}^{-1}\steeringvector$ (with $\steeringvector\in\mathbb{C}^N$ the steering vector) that defines the beamformer output is obtained as the expectation value of the QSVT unitary through a Hadamard test~\cite{NielsenChuang2010}, offering the ability to read the tomographic power profile out directly from the quantum computer.

Moreover, the QSVT formulation provides a joint control mechanism of circuit cost and spectral response. It acts as a polynomial spectral regularization controlled by its inverse approximation bound. Specifically, the condition number bound $\conditionthreshold$ acts selectively on the spectrum, faithfully inverting eigenvalues above $1/\conditionthreshold$ while suppressing those below, in contrast to the uniform shift of diagonal loading. It is closely related to the truncated singular value decomposition (SVD) and Tikhonov regularization~\cite{Hansen1987}, differing in that the bound arises intrinsically from the inversion rather than as an imposed prior. Lowering it moves the reconstruction away from the adaptive Capon response toward the non-adaptive Fourier one, and past it at the smallest bound we test.

The principal contributions of this work are as follows:
\begin{itemize}
    \item An end-to-end QSVT-based integration of the quantum Capon beamformer within a TomoSAR image processing chain, applied to real ESA AfriSAR P-band data over the Mondah forest and validated against LVIS LiDAR ground truth.
    \item A quantitative demonstration that quantum Capon reproduces the classical Capon tomogram with high fidelity for $\conditionthreshold \ge 148$ (aggregate peak-height RMSE against LiDAR and $\textrm{SSIM} >0.99$ at every track count).
    \item A straightforward formulation of the Capon power estimate, in which the beamformer denominator $\steeringvector^{\dagger}\mtrx{R}^{-1}\steeringvector$ is read out directly from the QSVT block encoding via a Hadamard test, so that only a single scalar per look elevation need be measured rather than the full inverse matrix.
    \item Identification and demonstration of the QSVT condition number bound as a spectrally selective regularizer for adaptive beamforming, qualitatively distinct from diagonal loading, and tunable to the requirements of the downstream retrieval task.
\end{itemize}

Finally, we highlight that the aim of this work is to introduce the SAR/image-processing community to quantum linear system solvers' (QLSS) methods than a demonstration of a speedup.
In this work all logarithms are base 2 unless otherwise specified.

\section{TomoSAR Spectral Estimation and the Capon Beamformer Spectrum Problem}
\label{sec:background}

\subsection{Tomographic signal model}

TomoSAR reconstructs the vertical distribution of backscattered power within each azimuth-range resolution cell from a stack of $N$ co-registered SAR acquisitions taken from slightly displaced orbital or flight tracks~\cite{Reigber2000, Tebaldini2012}. After phase calibration and de-ramping, the $N$ complex observations associated with a given resolution cell are collected into a data vector $\vctr{y}\in\mathbb{C}^{N}$.

In the continuous formulation, each acquisition measures the complex reflectivity profile $\gamma(s)$ along the height $s$ above the reference surface, modulated by a phase term set by that track's vertical wavenumber $\spatialfreq{z}{m}$~\cite{Franceschetti2018}:
\begin{equation}
  \label{eq:tomo-integral}
  y_m = \int \gamma(s)\,e^{\,\imagi \spatialfreq{z}{m} s}\,\mathrm{d}s,
  \qquad m = 1,\dots,N.
\end{equation}
Here $\spatialfreq{z}{m}$ is the wavenumber of the $m$-th track along the vertical direction $\vctr{z}$, determined by the acquisition geometry (baseline, look angle, slant range, and wavelength). The set $\{\spatialfreq{z}{m}\}_{m=1}^{N}$ fixes the elevation point-spread function and hence the achievable resolution. Specifically, a wider spread of vertical wavenumbers (a longer effective baseline aperture) yields finer elevation discrimination. Each observation is thus a sample, at spatial frequency $\spatialfreq{z}{m}$, of the Fourier transform of the elevation reflectivity, and the acquisition geometry determines which frequencies (i.e.\ which $\spatialfreq{z}{m}$) are observed. Recovering $\gamma(s)$, or its power $\mathcal{P}(s)=\mathbb{E}\,|\gamma(s)|^2$, from the irregular $\{\spatialfreq{z}{m}\}$ is the tomographic reconstruction problem.
 
Under the standard discrete-scatterer model, the contribution of a scatterer at elevation $s$ is described by the steering vector $\steeringvector(s)\in\mathbb{C}^{N}$:
\begin{equation}
  \label{eq:steering}
  \steeringvector(s) = \bigl[\,e^{\,\imagi k_{z,1} s},\; e^{\,\imagi k_{z,2} s},\; \dots,\; e^{\,\imagi k_{z,N} s}\,\bigr]^{\!\top},
\end{equation}
so that the exponent $e^{\,\imagi k_{z,m}s}$ of \eqref{eq:tomo-integral} becomes the \mbox{$n$-th} entry of $\vctr{a}(s)$ and the forward model is expressed as $\displaystyle\vctr{y}=\int \gamma(s)\,\vctr{a}(s)\,\mathrm{d}s$. The tomographic reconstruction problem is the estimation of the power $\mathcal{P}(s)$ as a function of height $s$ over a height axis spanning the ground and canopy returns, and the locations of its dominant peaks identify the physical scattering layers.

\subsection{Fourier and Capon estimators}
\label{sec:capon-est}

The conventional Fourier (beamforming) estimator forms the matched-filter output:
\begin{equation} \mathcal{P}_{\mathrm{Fourier}}(s) = \bigl|\,\vctr{a}(s)^\dagger\,\vctr{y}\,\bigr|^2, \label{eq:fourier}\end{equation}
which is non-adaptive, since its resolution is set entirely by the aperture through $\{k_{z,n}\}$ and cannot be improved by the data. When the ground and canopy returns fall within one Rayleigh elevation cell, as is common at the track counts available in practice, the Fourier estimator merges them into a single unresolved lobe.

The Capon, or Minimum Variance Distortionless Response (MVDR), estimator~\cite{Capon1969, LombardiniReigber2003} overcomes this limit by adapting to the second-order statistics of the data. It designs, for each height $s$, a weight vector $\vctr{w}(s)$ that minimizes the total output power subject to a unit-gain constraint in the look direction:
\begin{equation} \label{eq:mvdr-opt}
\min_{\vctr{w}}\;\vctr{w}^\dagger \mtrx{R}\,\vctr{w} \quad \text{subject to} \quad \vctr{w}^\dagger \vctr{a}(s)=1, 
\end{equation}
where the components of the covariance matrix $\mtrx{R} \in \Complex^{N \times N}$,
\begin{equation}\label{eq:covmat}
R_{ij} = \frac{1}{L}\sum_{l=1}^L y_i^{(l)} y_j^{{(l)}*},
\end{equation}
are estimated as a sample covariance over $L$ statistically homogeneous looks  $\{ \vctr{y}^{(l)} \}_l$ within a multi-look window. The closed-form solution of \eqref{eq:mvdr-opt} is:
\begin{equation} \label{eq:weights} 
\vctr{w}(s) = \mtrx{R}^{-1}\vctr{a}(s)\big/\bigl(\vctr{a}(s)^\dagger \mtrx{R}^{-1}\vctr{a}(s)\bigr),
\end{equation}
and the resulting power spectrum is:
\begin{equation}
    \mathcal{P}_{\mathrm{Capon}}(s)=\frac{1}{\vctr{a}(s)^\dagger\,\mtrx{R}^{-1}\,\vctr{a}(s)}.
    \label{eq:capon}
\end{equation}
The adaptive nulling encoded in $\mtrx{R}^{-1}$ suppresses power arriving from heights other than $s$, sharpening the elevation PSF below the Fourier limit and allowing the separation of ground and canopy returns that share a resolution cell. 

\section{Computing the Capon estimator}
\label{sec:classical-capon}

\subsection{Problem statement}
\label{sec:problem}
The computational task under study in this paper is, given
inputs $(\mtrx{M}, \vctr{v}) \in \Complex^{N\times N} \times \Complex^N$,
compute 
$\Re(\vctr{v}^{\dagger}\mtrx{M}^{-1}\vctr{v})$. 
When $\mtrx{M} = \covarianceM$ is Hermitian positive definite (HPD) with components \eqref{eq:covmat}, 
and $\vctr{v} = \steeringvectorS$ has the form \eqref{eq:steering} for some $s \in \Reals$, 
we are computing the 
denominator from~\eqref{eq:capon}, which we refer to as the Capon denominator problem (\Problem{}).

\subsection{The inversion bottleneck and diagonal loading}
\label{sec:bottleneck}

Equation~\eqref{eq:capon} shows that the entire adaptive behavior of the Capon estimator is carried by the inverse covariance matrix $\mtrx{R}^{-1}$. The covariance matrix is HPD by construction. Although the sample covariance is well estimated here (each multi-look cell averages several hundred looks, so $L \gg N$), $\mtrx{R}$ is frequently ill-conditioned, with a condition number $\kappa = \lambda_{\max}(\mtrx{R})/\lambda_{\min}(\mtrx{R})$. The conditioning is structural rather than from finite sampling, since the backscatter is dominated by a few bright scattering mechanisms (ground and canopy), which concentrate the signal energy into a small number of large eigenvalues, while closely spaced elevation returns render their steering vectors nearly collinear. Direct inversion amplifies the resulting wide spread of eigenvalues along the poorly constrained small eigenvalue directions, motivating the regularization of an intrinsically ill-posed inverse~\cite{Hansen1987}.

The standard fix is diagonal loading, in which a scaled identity is added to the covariance prior to inversion:
\begin{equation}
\mtrx{R}_{\ell} = \mtrx{R} + \ell\,\frac{\operatorname{tr}(\mtrx{R})}{N}\,\mtrx{I}, \label{eq:diagload}
\end{equation}
with loading factor $\ell>0$. Diagonal loading shifts \emph{every} eigenvalue of $\mtrx{R}$ by the same amount, improving the condition number and suppressing noise amplification at the cost of biasing the covariance towards a scaled identity and hence dampening the adaptive resolution. The factor $\ell$ thus controls a single, spectrally uniform trade-off between resolution and stability.

\subsection{Classical Resources}
\Problem{} can be computed in a straight forward way using standard classical (i.e\@. non-quantum) algorithms. 
A purely classical computer computes the diagonal loading factor
$\ell\trace{R}/N$  (see~\eqref{eq:diagload}) in $\bigO{N}$ arithmetic operations, 
inverts the matrix $\covarianceM$ in $\bigO{N^3}$ time,
and computes the inner product in $\bigO{N^2}$ operations.
The matrix inversion step dominates the run-time giving $\bigO{N^3}$ for the whole computation. 
The non-input memory required for this step is naively on the order of $\bigO{N^2}$.

\subsection{Approximate \Problem}
\label{sec:quant_problem}

In Section~\ref{sec:method} we describe how to compute an approximation of \Problem{} using a quantum computer.
The quantum algorithm presented requires two additional input values, 
the  bound 
$\conditionthreshold \in \Reals_+$
on $\covarianceM$ condition number
and approximation accuracy $\varepsilon \in \Reals_+$. 

For QSVT the requirement is to have a large enough value $\conditionthreshold$, 
which can be estimated using a number of methods~\cite{Martyn2021}. 
Since this $\conditionthreshold$ does not strictly depend on input $\covarianceM$ 
it is typically considered an input for this family of quantum algorithms~\cite{Martyn2021}.

The input $\varepsilon \in \Reals_+$ is the approximation error such that  $\steeringvectorS^{\dagger}P({\covarianceM})\steeringvectorS\in\mathbb{R}_+$ for the polynomial function $P$ on the eigenvalues of $\covarianceM$ in $[-1, -1/\conditionthreshold] \cup [1/\conditionthreshold,1]$, satisfying $\left|\steeringvectorS^{\dagger}\left(P(\covarianceM)-{\covarianceM}^{-1}\right)\steeringvectorS\right|<\varepsilon$.

With these extra parameters we refer to the Approximate \Problem{} problem.

\section{QSVT-Based Capon Inversion and Readout}
\label{sec:method}

An introduction to quantum computing is beyond the scope of this work, we assume that the reader is familiar with the basics of quantum computing~\cite{NielsenChuang2010}.

Our quantum implementation
proceeds in two stages. First, the inverse is realized as a QSVT (Sections~\ref{sec:be}--\ref{sec:circuit}). Second, rather than constructing $\mtrx{R}^{-1}$ explicitly, we obtain the Capon denominator directly as the expectation value of the QSVT unitary through a Hadamard test (Section~\ref{sec:readout}), giving an estimate 
of the tomographic power spectrum,  directly from the quantum circuit.
The circuit depth and approximation error are discussed in Section~\ref{sec:depth}.

\subsection{Block encoding of the covariance matrix}
\label{sec:be}

First, we note that since the covariance matrix $\mtrx{R}$ is a HPD matrix, its singular values are its eigenvalues.
To block-encode $\mtrx{R}$, we first need to normalize it:
\begin{equation*} 
\widetilde{\mtrx{R}} = \mtrx{R} / \alpha, 
\label{eq:normalize}
\end{equation*}
such that every eigenvalue $x_k = \lambda_k/\alpha \in (0,1]$. 
For a HPD matrix a straightforward choice is $\alpha = \Tr(\mtrx{R})$, since $\lambda_{\max}\leq \Tr(\mtrx{R})$ holds for any HPD matrix. However it is not the only choice. For example, one can choose $\alpha = \sum_{i} |a_i|$, where $a_i \in \Reals$ are coefficients in the Pauli decomposition of~$\mtrx{R}$, which upper bounds $\|\mtrx{R}\|_2$ by the triangle inequality. In each case, $\alpha/\lambda_{\max}\geq1$ inflates the condition number bound and hence the polynomial degree by the same factor.

A block encoding of $\widetilde{\mtrx{R}}$ is a unitary 
$U_{\widetilde{\mtrx{R}}}$ on $n+q$ qubits, with $n = \lceil \log N \rceil$ system qubits and $q$ ancilla qubits, whose projection on a (usually) zero subspace of ancilla qubits 
reproduces $\widetilde{\mtrx{R}}$:
\begin{equation*} 
\bigl(\mathbb{I}_{\mathrm{sys}}\otimes \bra{0}
_{\mathrm{anc}}\bigr)\, U_{\widetilde{\mtrx{R}}}\,\bigl(\mathbb{I}
_{\mathrm{sys}}\otimes \ket{0}_{\mathrm{anc}}\bigr) = \widetilde{\mtrx{R}}
\label{eq:be-def}.
\end{equation*}
In the simplest case, one can have $q=1$~\cite{Martyn2021}:
\begin{equation}\label{eq:be-r}
    U_{\widetilde{\mtrx{R}}} = 
    \begin{pmatrix}
        \widetilde{\mtrx{R}} & \mtrx{C} \\
        \mtrx{C} & -\widetilde{\mtrx{R}}
    \end{pmatrix}
    \!\cong\!
    \bigoplus_k \begin{pmatrix}
        x_k & \sqrt{1-x_k^2} \\
        \sqrt{1-x_k^2} & -x_k
      \end{pmatrix},
\end{equation}
where the matrix $C$ is the matrix $\widetilde{\mtrx{R}}$ with the function $f(x)=\sqrt{1-x^2}$ applied to its eigenvalues,
the symbol $\cong$ is an equivalence up to the basis, $\bigoplus_k$ is a direct sum over 
signal reflection operators $\mtrx{W}_R(x_k)=x_k\,\mtrx{Z}+\sqrt{1-x_k^2}\,\mtrx{X}$, where $Z$ and $X$ are the Pauli $Z$ and $X$ matrices, respectively.
The encoding is unitary because $\widetilde{\mtrx{R}}^2+\mtrx{C}^2=\mtrx{I}$. 
This encoding is usually assumed within QSVT, but
this is not the only encoding, that can be used~\cite{Gilyen2019, Martyn2021}. 

The block-encoding~\eqref{eq:be-r} 
is easy to understand and manipulate mathematically, but its compilation into hardware native gates is not always straightforward.
There are other block-encoding methods that are easier to implement, for example,
Fast Approximate Block Encoding 
(FABLE)~\cite{campsfable2022}, which encodes $\widetilde{\mtrx{R}}$ using $q = \lceil \log N \rceil + 1$ ancilla qubits, without diagonalizing $\widetilde{\mtrx{R}}$, and with error $\varepsilon_{be} \sim \bigO{N^3 \delta_c}$, where $\delta_c$ is a compression threshold. 
The gate complexity for this method is $\bigO{N^2}$, where $N^2$ CNOTs and $N^2$ single-qubit rotations are required.
There are also sparse variants that significantly reduce its gate count~\cite{kuklinski2024,Yang2025}. Another block-encoding method is the linear combination of unitaries (LCU), which exploits the Pauli decomposition of $\widetilde{\mtrx{R}}$ and realizes its block encoding via \textsc{prepare} and \textsc{select} oracles~\cite{Childs2012, Gilyen2019}. Alternatively, classical data stored in a quantum read-only memory (QROM)-friendly~\cite{babbush2018} data structure admit a block encoding with subnormalization $\|\widetilde{\mtrx{R}}\|_F$~\cite{kerenidis2017, chakraborty2019, clader2022}. Finally, if $\widetilde{\mtrx{R}}$ has exploitable structure, there are explicit circuits for sparse and structured matrices~\cite{Camps2024, Snderhauf2024} and hierarchical matrix encodings of dense kernels~\cite{Nguyen2022} give more efficient encodings.

\subsection{QSVT circuits and post-selection}
\label{sec:circuit}

First, we define a projector-controlled phase shift operator $\mtrx{S}_z(\phi)$, a key component of QSVT, which applies a phase rotation $R_z(-2\phi) = e^{\imagi \phi \sigma_z}$ 
to each signal operator $\mtrx{W}_R(x_k)$ in the block encoding~\eqref{eq:be-r}:
\begin{equation} \label{eq:phase-shift} 
\mtrx{S}_z(\phi) = e^{\imagi \phi (2\Pi - \mtrx{I})} 
= \bigoplus_k e^{\imagi \phi \sigma_z}, 
\end{equation}
where $\Pi = \mtrx{I}_{\mathrm{sys}} \otimes \ket{0}_{\mathrm{anc}}\bra{0}_{\mathrm{anc}}$ is a projector onto a zero ancilla subspace. 

By interleaving $d$ applications of the block encoding with $\mtrx{S}_z$
over a sequence of angles $\vctr{\phi} = \{ \phi_k \}_{k=1}^d$, QSVT applies a polynomial $\polyP$
to each of the singular values of a matrix block-encoded in a larger unitary $U_{\widetilde{\mtrx{R}}}$, which satisfies the following conditions~\cite{Martyn2021, Gilyen2019}:
\begin{enumerate}
    \item $\mathrm{deg}(\polyP) \leq d$,
    \item $\polyP$ has parity $d \mod 2$,
    \item $|\polyP| \leq 1$.
\end{enumerate} 
Because of the block encoding, the arguments of the polynomial within QSVT are always in the range $[-1,1]$.
The angles $\vctr{\phi}$ are dependent only on the type of the polynomial one wants to implement but not on the block-encoding size, and can be found by known techniques~\cite{Martyn2021}.
For odd $d$ the QSVT operator takes the form:
\begin{equation} 
U_{QSVT} = \mtrx{S}_z(\phi_1) U_{\widetilde{\mtrx{R}}} \left[\prod_{k=1}^{(d-1)/2}  \mtrx{S}_z(\phi_{2k}) U^{\dagger}_{\widetilde{\mtrx{R}}}\, \mtrx{S}_z(\phi_{2k+1}) U_{\widetilde{\mtrx{R}}}\right].
\label{eq:qsvt-seq}
\end{equation}
Projecting (post-selecting) onto a $\ket{0}_{\mathrm{anc}}$ subspace of the ancilla qubit subsystem~\cite{Martyn2021} gives the desired result:
\begin{equation}\label{eq:qsvt-seq-post-sel}
    \polyP(\widetilde{\mtrx{R}}) = (\mtrx{I}_\mathrm{sys}\otimes\bra{0}_\mathrm{anc}) U_{QSVT} (\mtrx{I}_\mathrm{sys}\otimes\ket{0}_\mathrm{anc}).
\end{equation}

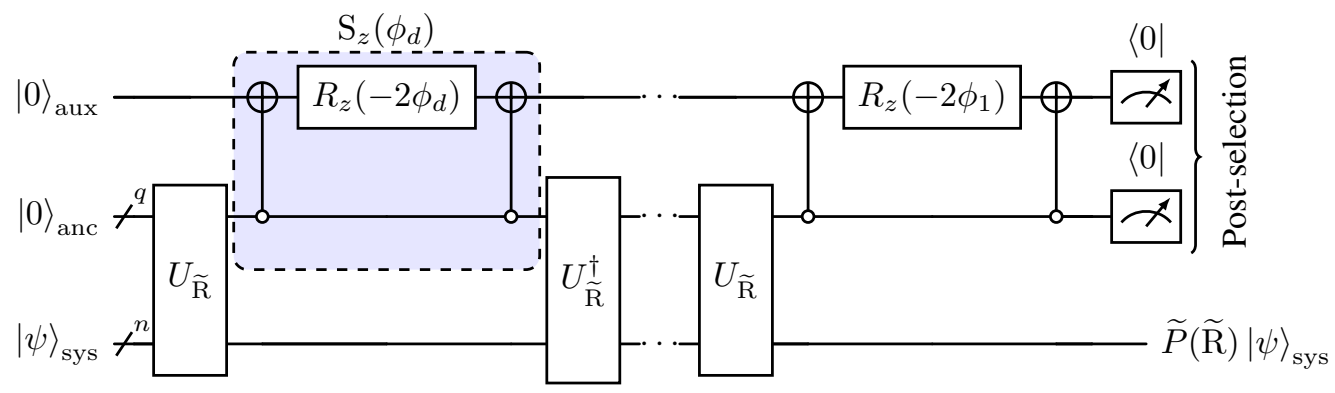
\begin{figure}[tb]
  \centering
  \begin{tikzpicture}
  \node[scale=0.65] {
  \begin{quantikz}[column sep=0.2cm]
    \lstick{$\ket{0}_{\mathrm{aux}}$} & \qw  &  \qw & \targ{}\gategroup[2,steps=3,style={dashed, rounded corners, inner sep=0.5pt,fill=blue!10},background]{$\mtrx{S}_z(\phi_d)$} & \gate{R_z(-2\phi_{d})} & \targ{} & \qw & \push{\cdots} & \qw & \targ{} & \gate{R_z(-2\phi_{1})} & \targ{} & \qw & \meter{\bra{0}} \rstick[2,label style={rotate=90, xshift=-30, yshift=-30}]{Post-selection} \\
    \lstick{$\ket{0}_{\mathrm{anc}}$} & \qwbundle{q} & \gate[2]{U_{\widetilde{\mtrx {R}}}} & \octrl{-1} & & \octrl{-1} & \gate[2] {U^\dagger_{\widetilde{\mtrx {R}}}} & \push{\cdots} & \gate[2]{U_{\widetilde{\mtrx  {R}}}} & \octrl{-1} & & \octrl{-1} & \qw & \meter{\bra{0}}  \\
    \lstick{$\ket{\psi}_{\mathrm{sys}}$} & \qwbundle{n} & \qw & & \qw && & \push{\cdots} & & \qw &&& \qw & \rstick{$\polyP(\widetilde{\mtrx{R}})\ket{\psi}_{\mathrm{sys}}$}
  \end{quantikz}
  };
  \end{tikzpicture}
  \caption{QSVT circuit implementation of $U_{QSVT}$ in Eq.~\eqref{eq:qsvt-seq} for the odd-degree $d$ polynomial $\widetilde{P}$ applied to the matrix $\widetilde{\mtrx{R}}$ block-encoded with $U_{\widetilde{\mtrx {R}}}$ and the post-selection procedure in Eq.~\eqref{eq:qsvt-seq-post-sel}. 
    }
  \label{fig:qsvt_seq}
\end{figure}

To implement the QSVT sequence on a quantum device, we need one auxiliary qubit for $\mathrm{S}_z(\phi)$ implementation. The full sequence is schematically shown in Fig.~\ref{fig:qsvt_seq}.

Matrix inversion is a special case, in which the applied polynomial $\polyP$ approximates $f(x)=x^{-1}$. Since $x^{-1}$ is singular at the origin, the approximation is restricted at the spectral interval $[1/\conditionthreshold,1]$ 
and is realized by a bounded polynomial $\widetilde{P}(x)$ of odd degree $d$ that satisfies:
\begin{equation*}
    \left| \Re( \widetilde{P}(x) ) - \frac{b}{x} \right| \leq \varepsilon, \quad \forall x\in[-1,-1/\conditionthreshold]\cup[1/\conditionthreshold,1] \label{eq:poly-acc},
\end{equation*}
where the scaling factor $b>0$ is fixed by condition 3) on $\polyP$ discussed above, $\varepsilon>0$ is approximation accuracy, and $\Re$ stands for the real part. 
We follow the procedure discussed in~\cite{Martyn2021} to obtain the angles $\vctr{\phi}$ to implement $\polyP$, where $d$ depends only upon $\conditionthreshold$ and $\varepsilon$ and scales as $\bigO{\conditionthreshold\log(\conditionthreshold/
\varepsilon)}$ (for more information on this see~\cite{Martyn2021, Gilyen2019}). Complete coverage of the normalized covariance spectrum requires 
$\conditionthreshold\ge\frac{\alpha}{\lambda_{\min}(\mtrx{R})}$. This equals the condition number only when $\alpha=\lambda_{\max}(\mtrx{R})$.
Applying $\widetilde{P}$ to the block encoded matrix gives an operator approximating $b\,\widetilde{\mtrx{R}}^{-1}$ with respect to the same tolerance:
\begin{equation*}
\bigl\|\Re(\polyP(\widetilde{\mtrx{R}})) - b\,\alpha\,
\mtrx{R}^{-1}\bigr\| \le \varepsilon. 
\label{eq:inv-bound}
\end{equation*}

To implement the calculation of the (unscaled) approximation $\steeringvector(s)^{\dagger} \Re(\polyP(\covarianceMscaled)) \steeringvector(s)$ of the looked-for scalar product, we will use a procedure known as the Hadamard test.

\subsection{Hadamard test estimation of the Capon denominator}
\label{sec:readout}

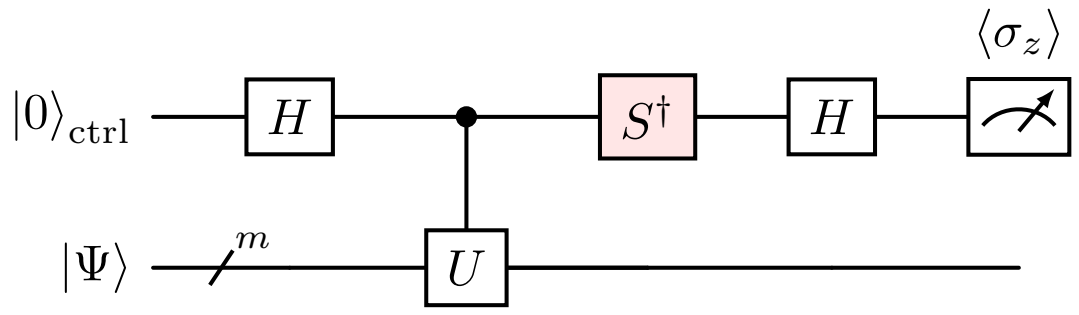
\begin{figure}[tb]
  \centering
  \begin{tikzpicture}
  \node[scale=0.95] {
  \begin{quantikz}[column sep=0.65cm, row sep=0.5cm]
    \lstick{$\ket{0}_{\mathrm{ctrl}}$} & \gate{H} & \ctrl{1} & \gate[style={fill=red!10}]{S^\dagger} & \gate{H} & \meter{\langle \sigma_z \rangle} \\
    \lstick{$\ket{\Psi}$}  & \qwbundle{m} & \gate{U} & \qw  &  &
  \end{quantikz}
  };
  \end{tikzpicture}
  \caption{Hadamard test estimation~\cite{NielsenChuang2010} of the expectation value $\Re \bra{\Psi}U\ket{\Psi} = \bra{\Psi}(U+U^\dagger)\ket{\Psi}/2$ (circuit without the $S^\dagger$ gate in red) and $\Im \bra{\Psi}U\ket{\psi} = \bra{\Psi}(U-U^\dagger)\ket{\Psi}/(2i)$ (circuit with the $S^\dagger$ gate in red) by measuring the expectation value $\langle \sigma_z \rangle$ on the control qubit only. }
  \label{fig:hadamard-test}
\end{figure}

Given an $m$-qubit state $\ket{\Psi}$ as a starting state of a quantum system and a unitary $U$, the Hadamard test procedure~\cite{NielsenChuang2010} allows for estimating the real and imaginary parts of a scalar product $\bra{\Psi}U\ket{\Psi}$, by introducing and measuring one additional control ancilla qubit. The Hadamard test procedure can be performed in the following steps:
\begin{enumerate}
  \item Introduce an additional control qubit, initialised in the state $\ket{+}_\mathrm{ctrl} = H\ket{0}_\mathrm{ctrl}$.
    The other qubits (referred to as data qubits in what follows)
    are initialised in the ground state $\ket{\Psi}$.
  \item Perform a unitary operation $U$ on the data qubits,
    controlled by the control qubit.
  \item[3.1)] To estimate $\Re \bra{\Psi}U\ket{\Psi} = \bra{\Psi}(U+U^\dagger)\ket{\Psi}/2$,  apply a Hadamard gate on the control qubit, and measure the expectation value $\langle \sigma_z \rangle$ on it. 
  \item[3.2)] To estimate $\Im \bra{\Psi}U\ket{\Psi} = \bra{\Psi}(U-U^\dagger)\ket{\Psi}/(2i)$, apply the $S^\dagger$ gate and Hadamard on the control qubit, and measure the expectation value $\langle \sigma_z \rangle$ on it.
\end{enumerate}
The circuit for the Hadamard test is given in Fig.~\ref{fig:hadamard-test}.  

\begin{figure}[tb]
  \centering
  \begin{tikzpicture}
  \node[scale=0.95] {
  \begin{quantikz}[column sep=0.65cm, row sep=0.5cm]
    \lstick{$\ket{0}_{\mathrm{ctrl}}$} & \gate{H} & \ctrl{1} &  \gate{H} & \meter{\langle \sigma_z \rangle} \\
    \lstick{$\ket{0}_{\mathrm{aux}}$}  & \qw & \gate[3]{U_{QSVT}} & \qw  &  \\
    \lstick{$\ket{0}_{\mathrm{anc}}$}  & \qwbundle{q} &                      & \qw  &  \\
    \lstick{$\ket{a(s)}_{\mathrm{sys}}$}  & \qwbundle{n} &              & \qw & 
  \end{quantikz}
  };
  \end{tikzpicture}
  \caption{Hadamard test readout of the expectation value $\langle \sigma_z \rangle = \bra{a(s)}_{\mathrm{sys}}\Re(\polyP(\widetilde{\mtrx{R}}))\ket{a(s)}_{\mathrm{sys}}$ on the control qubit for estimating the Capon denominator $\steeringvector(s)^{\dagger}\mtrx{R}^{-1}\steeringvector(s)$. 
  }
  \label{fig:hadamard-test-qsvt}
\end{figure}
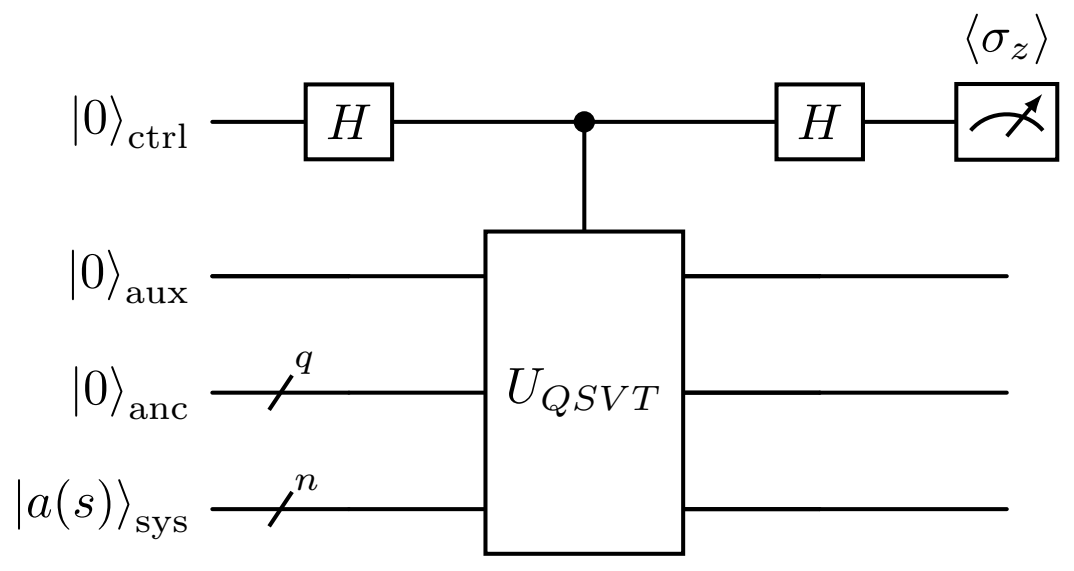

Using the Hadamard test with $U=U_{QSVT}$ and $\ket{\Psi} = \ket{a(s)}_{\mathrm{sys}} \otimes \ket{0}_{\mathrm{anc}} \otimes \ket{0}_{\mathrm{aux}}$, where $\ket{a(s)} = \steeringvectorS/\|\steeringvectorS\|$ and $\|\steeringvectorS\| = \sqrt{N}$, we can measure 
\[
    \Re \bra{\Psi} U_{QSVT} \ket{\Psi} = \bra{a(s)}_\mathrm{sys} \Re(\polyP(\widetilde{\mtrx{R}})) \ket{a(s)}_\mathrm{sys},
\]
which estimates $\steeringvectorS^\dagger\mtrx{R}^{-1}\steeringvectorS b\alpha/ N $ up to the accuracy $\varepsilon$ of the polynomial approximation of the inversion function and the statistical error $\delta$ due to the finite number of shots $M_S$, which scales as $\bigO{1/\sqrt{M_S}}$ according to Hoeffding's inequality. Using amplitude estimation techniques~\cite{Brassard2002}, we can improve this scaling to $\bigO{1/M_S}$.
The full circuit implementation of this is given in Fig.~\ref{fig:hadamard-test-qsvt}. 
The Capon denominator is estimated from the Hadamard-test  expectation. The auxiliary and block-encoding ancillae need not be post-selected during this scalar readout.

Preparation of the system qubits in a particular state is usually done using so-called quantum state preparation algorithms~\cite{Volya2024, Zhang2022}, which in the general case of $\ket{a(s)}$ can be computationally expensive.

\subsection{Quantum Resource Analysis}
\label{sec:depth}
We now analyze the quantum resources required to solve Approximate \Problem{}
(Section~\ref{sec:quant_problem}) via the algorithm outlined in 
Section~\ref{sec:readout}. 
Recall that the number of qubits is $n = \log_2 N$.

The first Approximate \Problem{} input is the vector $\steeringvectorS{} \in \Complex^N$ which must be 
amplitude encoded as a state $\ket{a(s)}$. 
Exact preparation of an arbitrary state $\ket{a(s)}$ for $s\neq 0$ requires $\bigO{N}$ gates, and
  depth $\bigO{N/\log N}$ (without ancillas)~\cite{Sun2023}.
However, cases where 
$\steeringvectorS$ has a special structure 
 i.e. such that their quantum representation has the form
  $\ket{a(s)} = \bigotimes_{p=1}^{M} \ket{a_{p}(s)}$ with $M \leq n$,
   we can use the general encoding algorithm for each of the $M$ subsystems,
    which can significantly reduce the cost of encoding.
In the limiting case of a fully separable state ($M=n$) or $s=0$,
the depth of encoding is constant.
Assuming the form $\ket{a_{p}(s)} = (e^{\imagi k'_{z,p}s}\ket{0} + e^{\imagi k'_{z,p+n}s} \ket{1})/\sqrt{2}$ 
for each $p$, full separability is equivalent to the requirement that the following system of $N$ equations
 is satisfied:
\begin{equation}\label{eq:wavenumber-cond}
    k_{z,m} = \sum_{p_1 \in \,O \setminus g(m)} \, k'_{z,p_1} + \sum_{p_2 \in \, g(m)} k'_{z,p_2+n} 
\end{equation}
where 
$O = \{ i \}_{i=1}^n$,  and 
$g(x) = \{i \in O \mid \lfloor \frac{x}{2^{i-1}}\rfloor \bmod 2 = 1 \}$ is the set of positions of the nonzero bits in the binary encoding of $x$.

The second input is the matrix $\covarianceMscaled\in \Complex^{N\times N}$,
 which must be block-encoded into a unitary $U_{\widetilde{\mtrx{R}}}$.
The cost of constructing the block-encoding circuit depends on the technique used;
for example FABLE requires $\bigO{N^2\log N}$ gates~\cite{campsfable2022},
 while LCU requires a unitary decomposition of $\covarianceMscaled$,
 (which is not always given) and requires $\bigO{N^2}$ gates~\cite{Low2019}. 

The 
structure and query complexity 
of the QSVT circuit is fixed for each pair $(\conditionthreshold, \varepsilon)$ which are specified as problem input.
The angles $\vctr{\phi}$ are
computed once per $(\conditionthreshold, \varepsilon)$.
Using the phase solver of~\cite{Dong2021} this requires
$\bigO{\conditionthreshold^2 \log^2(\conditionthreshold/\varepsilon)}$ classical arithmetic operations. 
The number of quantum gates in the QSVT circuit will scale as $\bigO{\conditionthreshold \log(\conditionthreshold/\varepsilon)}$~\cite{Martyn2021}.

The combined gate count for $U_{QSVT}$ (assuming FABLE block-encoding) is 
$\bigO{N + N^2 \log N \cdot \conditionthreshold \log(\conditionthreshold/\varepsilon)}$. 
Using the Hadamard test increases the number of gates by 2.

Measuring the Hadamard test result to an accuracy of $\delta$ 
requires $\bigO{1/\delta^2}$ shots; amplitude
estimation~\cite{Brassard2002} reduces this to $\bigO{1/\delta}$ (coherent)
repetitions at the cost of deeper circuits (Section~\ref{sec:readout}).

\section{Validation Methodology}
\label{sec:validation}

We simulate the quantum Capon beamformer to validate it according to two criteria.  The first criterion is numerical accuracy, that is, how closely does the quantum Capon beamformer reproduce the classical one. We use the metrics: peak height RMSE difference, structural-similarity (SSIM), and peak signal-to-noise ratios (PSNR).
The second criterion is physical validity, which compares the quantum Capon reconstructed scattering heights against the conventional Fourier beamformer (a non-adaptive reference), and 
an independent ground truth (measured with airborne LiDAR).
We compare results using peak height and center-of-gravity (CoG) statistics. The latter also shows the spectral selectivity of the QSVT bound (Section~\ref{sec:res_regularizer}). 
Both criteria are evaluated over a two-dimensional sweep in track count $N$ and condition number bound $\conditionthreshold$.
We further verify, at the statevector level on representative pixels, that the Hadamard-test readout of Section~\ref{sec:readout} reproduces the Capon denominator computed from the extracted QSVT inverse.

\subsection{Dataset and preprocessing}
\label{sec:dataset}
 
The test data are P-band single-look complex (SLC) acquisitions from the ESA AfriSAR campaign~\cite{Pardini2018} over the Mondah forest, Gabon, composed of $N_{\max}=11$ tracks with perpendicular baselines spanning $-80$ to $+\SI{80}{\metre}$. We use the HV polarization throughout our whole analysis. Each SLC is co-registered and flattened to the reference digital elevation model (DEM) measured with the F-SAR product (with the height axis set so that $s=0$ coincides with the processor's terrain reference) prior to tomographic focusing. The ground truth is acquired from co-located LiDAR products from the LVIS instrument~\cite{Blair1999}. Specifically, a CHM and a DTM, resampled to the tomographic grid are used. A constant \SI{4}{\metre} offset between the LVIS DTM and the DEM reference surface is removed before any height comparison.
 
The covariance matrix at each cell is estimated by multi-looking the SLC stack. At pixel spacing (\SI{0.81}{\metre} range, \SI{1.20}{\metre} azimuth) a target resolution of $20\times\SI{20}{\metre}$ corresponds to a $25\times17$ look window, so each sample covariance $\mtrx{R}$ is formed from up to $L=425$ looks (well above the $N\le 11$ tracks) so that $\mtrx{R}$ is well estimated and the ill-conditioning discussed in Section~\ref{sec:background} is not stemming from a finite sampling effect. 
Multi-looking on this grid yields approximately \num{12820} valid pixels, which is held fixed across every $(N,\conditionthreshold)$ configuration so that all comparisons are made on an identical pixel set. Height metrics are evaluated over the LiDAR-valid subset.
 
\subsection{Reconstruction and sweep parameters}
\label{sec:sweep}

For each pixel, the tomographic power profile is reconstructed on a height axis $s\in[-20,100]$~m above the reference DEM, sampled at $101$ points (\SI{1.2}{\meter} spacing), and the three beamformers share the same sample covariance and steering vectors. The Fourier estimate~\eqref{eq:fourier} requires no inversion and is formed from the unloaded covariance as $\steeringvectorS^\dagger\mtrx{R}\,\steeringvectorS$, while both the classical and quantum Capon invert the diagonally-loaded covariance matrix $\mtrx{R}_\ell$ from~\eqref{eq:diagload} with $\ell = 5\times 10^{-3}$. 
Applying the same diagonal loading to both adaptive methods ensures that any difference between them is attributable to the inversion algorithm only and not to a difference in conditioning. 
The quantum Capon tomograms are computed by simulating the QSVT circuit, extracting the approximate inverse from its block, and evaluating the output power $\mathcal{P}_{\mathrm{QSVT}}$ of the resulting Capon weights; since these weights are invariant under a global rescaling of the inverse, the normalization $b\alpha$ does not affect the tomogram. For $\conditionthreshold \ge \alpha/\lambda_{\min}(\mtrx{R}_\ell)$, which reduces to  $\conditionthreshold\ge\conditionnumber$ for the normalization $\alpha=\lambda_{\max}(\mtrx{R}_\ell)$ used in our simulations, this coincides, up to the polynomial error, with the Hadamard-test estimate of Section~\ref{sec:readout}.

The track count $N$, which controls the effective baseline aperture, is swept through the following set $\{2,3,\dots,11\}$, where for each $N$ the tracks are selected with approximately uniform baseline spacing across the sorted $11$-track range, so that increasing $N$ adds aperture without resampling the same baselines. Increasing $N$ also increases the covariance dimension and tends to raise its condition number for the dataset we have chosen. 
The condition number bound $\conditionthreshold$ is swept through the  $\{8, 15, 37, 55, 81, 119, 148, 197, 289, 362, 481\}$, each with a fixed approximation tolerance $\varepsilon=0.1$ and a corresponding QSP phase sequence precomputed once and reused across all pixels and track counts. 
The physical condition number $\kappa$ used to interpret the results is measured per pixel \emph{after} diagonal loading.

\subsection{Metrics}
\label{sec:metrics}

\paragraph{Peak-height statistics}
The peak height $\hat{s}_i$ at the $i$-th pixel is the height of the maximum backscattered power in the reconstructed profile. With $\Delta s_i=\hat{{s}}_i-{s}_i^{\mathrm{ref}}$ the error relative to the LiDAR reference ($\mathrm{ref}=\{\mathrm{CHM},\mathrm{DTM}\}$) over the $L$ valid pixels, we report the root-mean square error $\mathrm{RMSE} = \sqrt{L^{-1}\sum_{i=1}^L \Delta s_i^2}$, the average bias $L^{-1}\sum_{i=1}^L \Delta s_i$, and the standard deviation $\sigma(\Delta s_i)$. Peak height is compared primarily against the CHM (top-of-canopy), since the dominant return is the canopy scatterer.

\paragraph{Center-of-gravity statistics}
The center-of-gravity (CoG) height:
\begin{equation}
    \label{eq:cog}
    s_{\mathrm{CoG}} = \frac{\displaystyle\sum_{k=1}^L \mathcal{P}(s_k)\,s_k}{\displaystyle\sum_{k=1}^L \mathcal{P}(s_k)}
\end{equation}
is the power-weighted mean elevation of the profile $\mathcal{P}(s_k)$. Instead of reporting a single maximum, it integrates over the full profile, making the CoG sensitive to sidelobe structure, dynamic range, and spectral leakage (the features that distinguish adaptive from non-adaptive beamformers and, as Section~\ref{sec:results} shows, the features on which 
the QSVT regularizer acts). CoG bias and standard deviation are computed against both the CHM and the DTM, similarly to the peak-height statistics.
\\
\paragraph{QSVT-versus-classical similarity}
To quantify how faithfully the QSVT reconstruction reproduces the classical Capon one, independently of the LiDAR reference, we compare the two on decibel scaled horizontal slices extracted at reference heights $s_0\in\{0,10,20,30\}$\,m above the DTM. On each slice we compute the SSIM~\cite{Wang2004}, which captures luminance, contrast and structural agreement, and the $\mathrm{PSNR}=10\log_{10}(\Delta^2/\mathrm{MSE})$, where $\Delta$ is the dynamic range of the classical Capon slice and $\mathrm{MSE}=N_{\mathrm{pix}}^{-1} \sum_{i=1}^{N_\mathrm{pix}} \left(\mathcal{P}^{\mathrm{dB}}_{\mathrm{Capon},i}(s_0) - \mathcal{P}^{\mathrm{dB}}_\mathrm{QSVT,i}(s_0)\right)^2$ is the mean-squared error between the QSVT and classical slices. These two metrics, together with the peak-height RMSE difference, form the algorithmic-correctness axis; the peak-height and CoG statistics against LiDAR form the physical-validity axis.

\section{Simulation results}
\label{sec:results}

\subsection{Physical validity against LiDAR}
\label{sec:res_validity}

\begin{figure*}[!tb]
  \centering
  \begin{subfigure}{0.49\textwidth}
    \includegraphics[width=\linewidth]{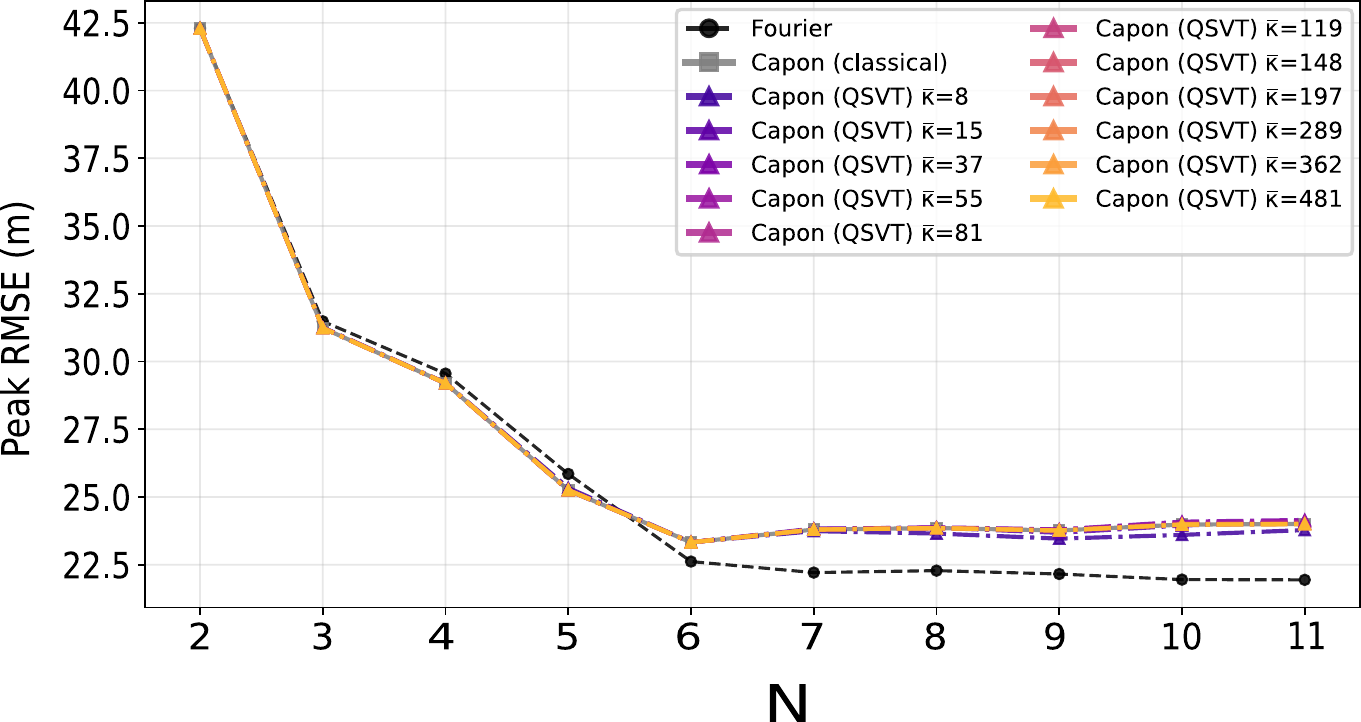}
    \caption{Peak-height RMSE vs CHM}
    \label{fig:rmse_vs_tracks}
  \end{subfigure}\hfill
  \begin{subfigure}{0.49\textwidth}
    \includegraphics[width=\linewidth]{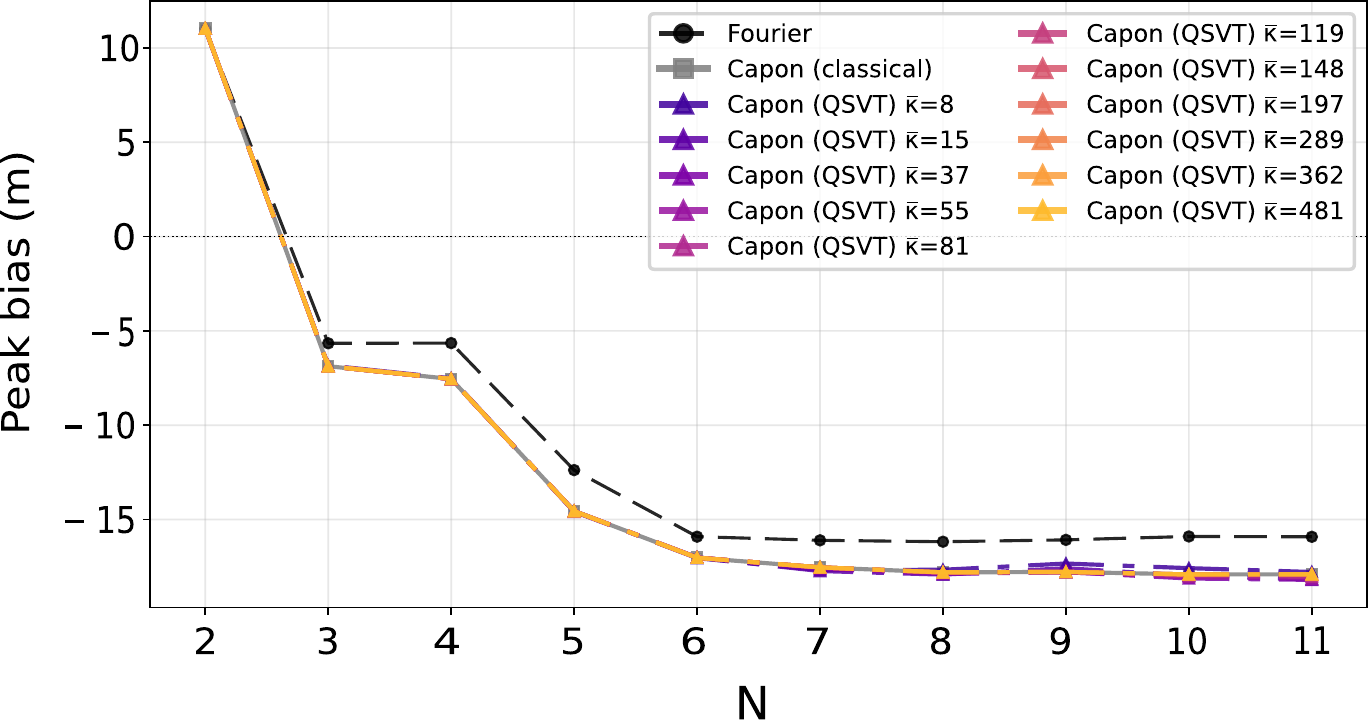}
    \caption{Peak-height bias vs CHM}
    \label{fig:bias_vs_tracks}
  \end{subfigure}
  \caption{Peak-height performance against the LiDAR canopy-height model as a function of track count $N$: (a) RMSE, (b) bias. Curves are shown for Fourier (dashed), classical Capon (grey), and quantum Capon at all tested condition number bounds $\conditionthreshold\in\{8,15,\dots,481\}$. 
  }
  \label{fig:rmse_bias_tracks}
\end{figure*}

Figure~\ref{fig:rmse_bias_tracks} shows the peak height error against the LiDAR CHM as a function of track count. All three beamformers exhibit the same behavior: the RMSE falls quickly from approximately $\SI{42.5}{\metre}$ at $N=2$ to a plateau of $22$--$\SI{24}{\metre}$ by $N\approx 6$, and the peak height bias moves from $+\SI{11}{\metre}$ at $N=2$ to a plateau for $N\ge 7$, near $-\SI{16}{\metre}$ for Fourier and between $-\SI{17.5}{\metre}$ and $-\SI{18}{\metre}$ for both Capon estimators. The plateau in RMSE coincides with the emergence of two-layer elevation resolution, since once the aperture is wide enough to separate the ground and canopy returns, the profile maximum locks onto the canopy. The residual error is set by the intrinsic offset between the P-band backscatter center and the top-of-canopy height reported by LVIS. This offset (the $\sim\!\SI{70}{\centi\metre}$ wavelength penetrates the canopy and scatters most dominantly off larger woody structure and the ground--trunk double bounce, locating the scattering centroid below the leaf surface) is the origin of both the $\sim\!\SI{22}{\metre}$ RMSE floor and the systematic negative bias, and it is shared by all three methods. The strongest P-band scattering elevation does not need to coincide with the LiDAR canopy top, so these errors should not be interpreted directly as canopy-height retrieval accuracy.

The two adaptive methods (classical and quantum Capon) track each other closely across the entire range, and the Fourier baseline is marginally apart. Its RMSE is slightly lower at high $N$ (a consequence of its broader profile placing the maximum fractionally higher, closer to the CHM) and its bias correspondingly slightly less negative. The key observation for the algorithmic comparison is that the QSVT curves have no visible dependence on $\conditionthreshold$ in any of the peak height metrics of Fig.~\ref{fig:rmse_bias_tracks}. The largest RMSE difference from classical Capon is $\SI{0.38}{\metre}$ (Fig.~\ref{fig:rmse_diff}). Hence, the choice of phase polynomial does not move the dominant peak, a point we return to in Section~\ref{sec:res_regularizer}.

\subsection{Algorithmic correctness against classical Capon}
\label{sec:res_correctness}

\begin{figure}[tb]
  \centering
  \includegraphics[width=\linewidth]{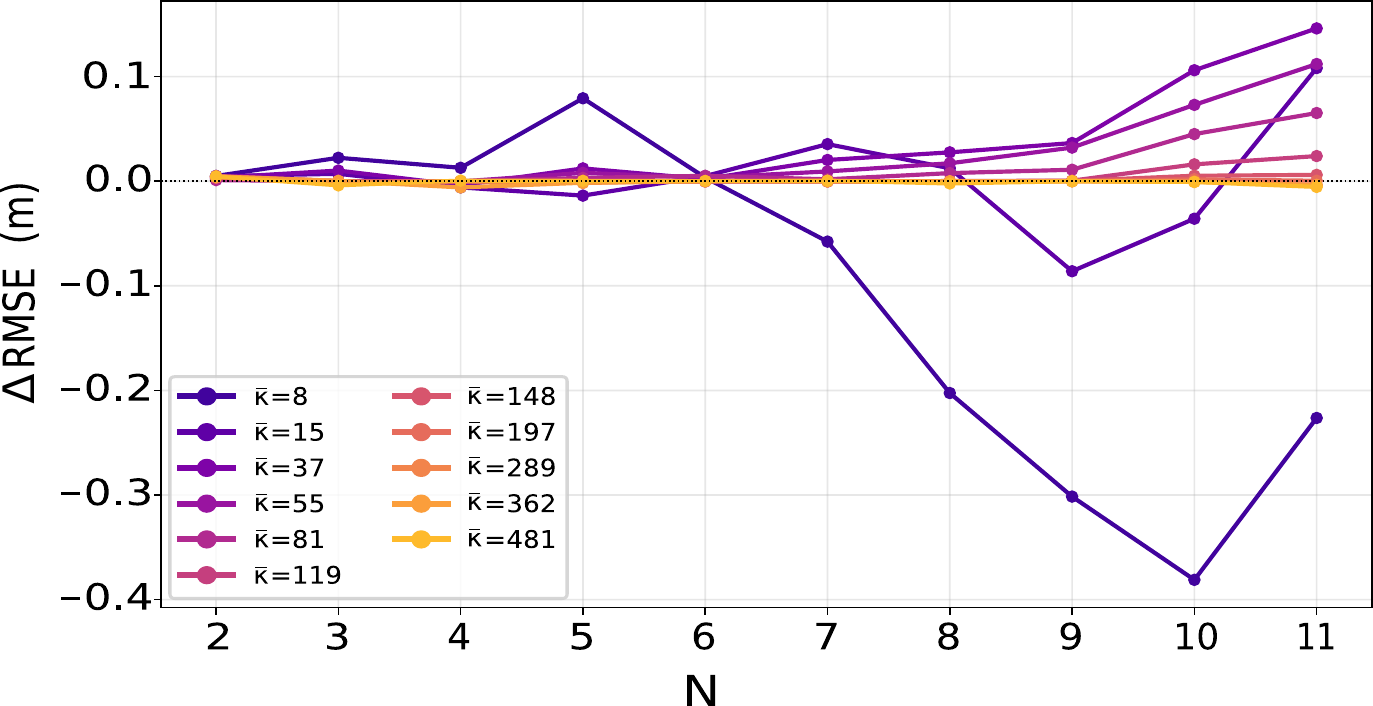}
  \caption{Difference in peak height RMSE between quantum and classical Capon versus $N$ and $\conditionthreshold$. 
  }
  \label{fig:rmse_diff}
\end{figure}

We quantify the inversion accuracy of QSVT by using three pixel-wise comparisons against classical Capon under identical diagonal loading. These are, the peak height RMSE difference, SSIM and PSNR ratios between the two reconstructions.
\\
\paragraph{Peak height RMSE difference}
Figure~\ref{fig:rmse_diff} shows the difference $\Delta\,\mathrm{RMSE}=\mathrm{RMSE}_{\mathrm{QSVT}}- \mathrm{RMSE}_{\mathrm{classical}}$ as a function of $N$ and $\conditionthreshold$. For $\conditionthreshold\ge 148$ the difference remains within $\pm \SI{0.02}{\metre}$ for all track counts. This confirms that the QSVT polynomial reproduces the classical inversion successfully whenever its settings match the pixel data. This is our primary algorithmic correctness result where at sufficient $\conditionthreshold$, the quantum inverter and exact inversion yield similar aggregate peak-height RMSE against LiDAR.

What is informative is that the difference fans out only at low $\conditionthreshold$ and high $N$. At $\conditionthreshold=8$ it is negative from $N=7$ on, reaching $\sim\!-\SI{0.38}{\metre}$ at $N=10$. For $\conditionthreshold=37$--$81$ it becomes positive, up to $\sim\!+\SI{0.14}{\metre}$ ($\conditionthreshold=37$, $N=11$), while $\conditionthreshold=15$ falls to $-\SI{0.09}{\metre}$ at $N=9$ before increasing to $+\SI{0.10}{\metre}$ at $N=11$. A very low bound suppresses the small eigenvalue contributions that Capon relies on for sidelobe nulling. The profile broadens and its power moves upward (Section~\ref{sec:res_regularizer}), which brings the dominant peak of some pixels closer to the CHM and lowers the RMSE at $\conditionthreshold=8$. The sign of $\Delta \mathrm{RMSE}$ does not however identify the regularization regime. For example, $\conditionthreshold=15$ has $\Delta \mathrm{RMSE}>0$ at $N=11$ although its CoG has already reached the Fourier value (Section~\ref{sec:res_regularizer}). All of these differences are small, at most $\SI{0.38}{\metre}$ against an RMSE of $22$--$\SI{24}{\metre}$ and a height sampling of $\SI{1.2}{\metre}$. We therefore use $\Delta \mathrm{RMSE}$ as a correctness check and the CoG to characterize the regularization effect.

\begin{figure}[tb]
  \centering
  \includegraphics[width=\linewidth]{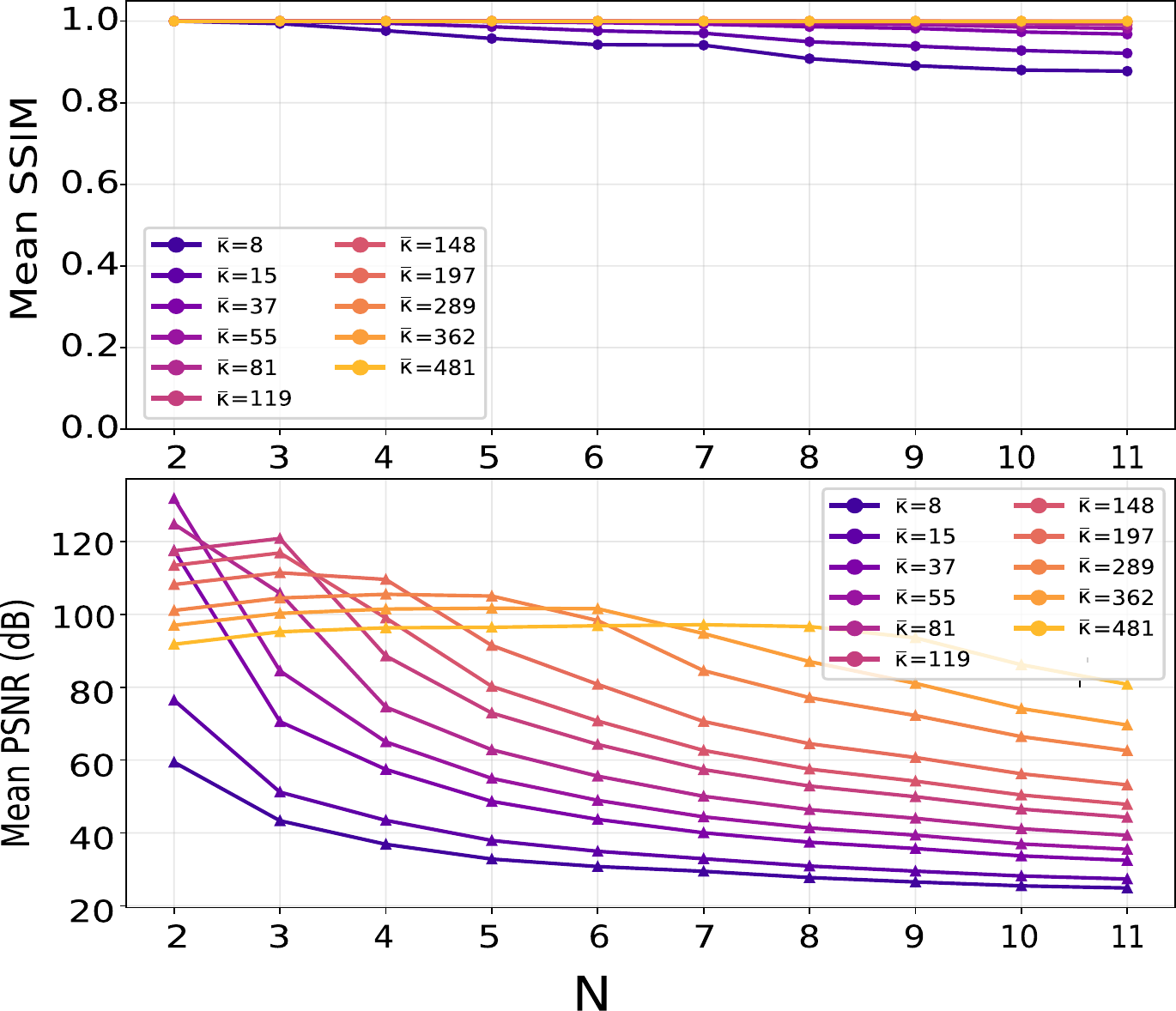}
  \caption{Mean SSIM (top) and mean PSNR (bottom) for quantum and classical Capon reconstructions
  averaged over four horizontal slices
  at $s_0\in\{0,10,20,30\}~$m above the DTM. 
  The x-axis is the track count $N$.
    }
  \label{fig:ssim_psnr}
\end{figure}

\paragraph{SSIM and PSNR}
Figure~\ref{fig:ssim_psnr} shows the mean SSIM and PSNR between the quantum and classical height slices. The SSIM approaches unity for the larger tested bounds across every track count, and is lower, $0.88$, at $\conditionthreshold=8$ and $N=11$. It is monotonic in $\conditionthreshold$ at fixed $N$ (higher bound, closer to classical) and decreases with $N$ at fixed $\conditionthreshold$, consistent with the spectrum of $\mtrx{R}$ spreading as the aperture size $N$ grows.

The PSNR depends non-monotonically on $\conditionthreshold$, and the best bound increases with $N$. At $N=2$ the highest PSNR, 131~dB, is reached at $\conditionthreshold=55$, while both the smallest bound ($\conditionthreshold=8$, 59~dB) and the largest ($\conditionthreshold=481$, 92~dB) sit lower. The best bound then rises to $\conditionthreshold=119$ at $N=3$, $197$ at $N=4$, $289$ at $N=5$, and $362$ at $N=6$. For $N\ge 7$ the ordering is monotonic in $\conditionthreshold$ and the PSNR decreases with decreasing bound, from 80~dB at $\conditionthreshold=481$ to 25~dB at $\conditionthreshold=8$ at $N=11$. The two regimes reflect the two binding constraints on quantum-versus-classical agreement. At small $N$ the spectrum of $\mtrx{R}$ is concentrated near unity, so most eigenvalues lie well inside the polynomial's well-approximated interval and the dominant disagreement is the cumulative numerical error of the degree-$d$ product, which grows with $\conditionthreshold$, penalizing the largest bounds. As $N$ grows and the spectrum spreads, an increasing fraction of eigenvalues approach or fall below $1/\conditionthreshold$. Hence, the polynomial's spectral coverage becomes binding, penalizing the smallest bounds. The optimal bound with respect to PSNR therefore tracks the spread of the spectrum. Even in the worst case ($\conditionthreshold=8,\,N=11$), a PSNR of $\sim\!\SI{25}{\decibel}$ corresponds to a root mean squared difference $\sim 6\%$ of the dynamic range, so the two reconstructions remain close.

The three correctness metrics show that for $\conditionthreshold\geq 148$ our QSVT implementation reproduces classical Capon to within $\SI{0.02}{\metre}$ in aggregate peak-height RMSE and above $0.99$ in mean SSIM at every track count. The only systematic errors occur at low $\conditionthreshold$. This is due to the regularization behavior, as we show below and not a failure of the inversion.

\subsection{Visual comparison of vertical profiles}
\label{sec:res_profiles}

\begin{figure}[tb]
  \centering
  \includegraphics[width=1.05\linewidth]{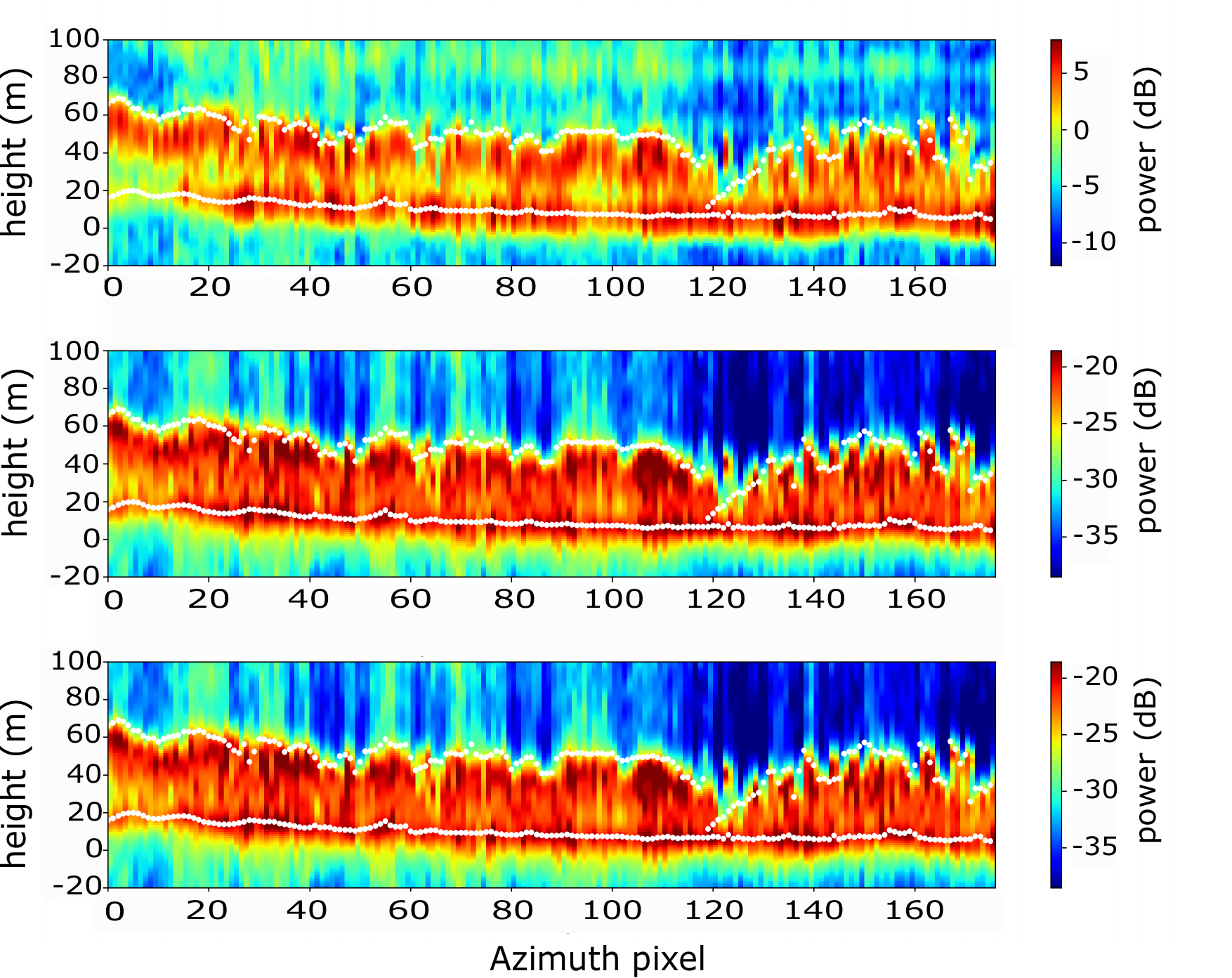}
  \caption{Vertical tomographic profiles along an azimuth line at fixed range, $N=11$, $\conditionthreshold=289$. Top: Fourier; middle: classical Capon; bottom: quantum Capon. Color encodes power in dB (note the differing scale of the Fourier panel); white dots mark the LiDAR DTM (lower) and DTM\,$+$\,CHM (upper) in the height frame of the tomograms. 
    }
  \label{fig:profiles_all}
\end{figure}

The quantitative agreement of Section~\ref{sec:res_correctness} can also be verified directly from the reconstructed tomograms. Figure~\ref{fig:profiles_all} shows the vertical power profiles along an azimuth line in fixed range for the three beamformers at $N=11$ and $\conditionthreshold=289$, with overlaid LiDAR DTM and DTM+CHM markers. The Capon panels (classical and quantum) put the energy between the LiDAR markers, with little inter-layer leakage. In contrast, the Fourier panel has the returns spread out and much of the power ends up smeared between layers. This is what the nulling looks like in the image itself. The CoG analysis below quantifies it. Additionally, the classical and quantum Capon panels are visually indistinguishable in band sharpness, dynamic range, and azimuth structure. They are consistent with the high SSIM and peak agreement of Section~\ref{sec:res_correctness}. Figure~\ref{fig:profiles_all} is thus a visual confirmation of the accuracy where at sufficient $\conditionthreshold$, the QSVT inversion produces a Capon tomogram equivalent to its classical counterpart.

\subsection{Selective regularization by the condition number bound $\conditionthreshold$}
\label{sec:res_regularizer}

The CoG metrics provide the clearest separation between methods and show the regularization behavior that distinguishes the QSVT inverter from its classical counterpart. Unlike the peak height, the CoG integrates over the full profile and is sensitive to sidelobe structure and dynamic range.
 
\begin{figure}[tb]
  \centering
    \includegraphics[width=\linewidth]{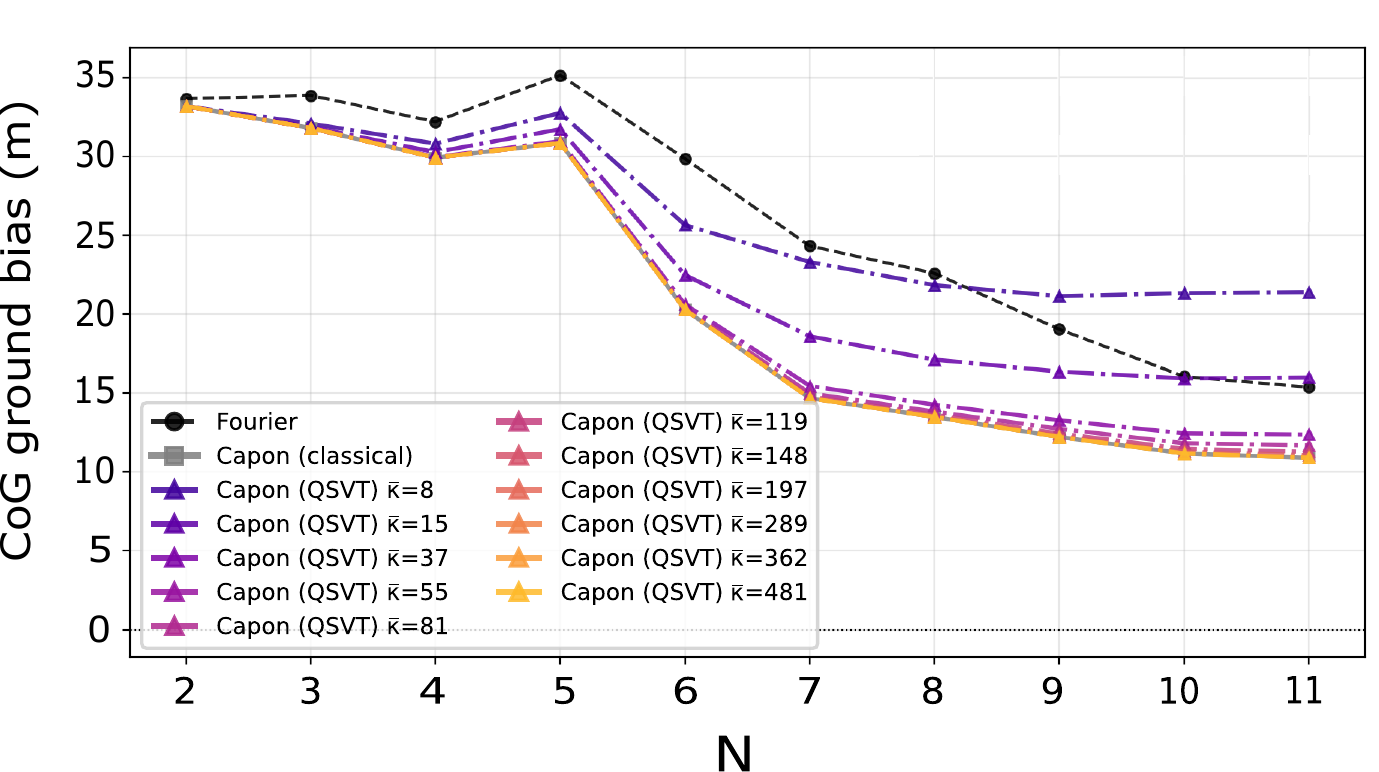}
    \label{fig:cog_dtm}
  \caption{CoG bias against the DTM, versus track count, for Fourier (dashed), classical Capon (grey), and quantum Capon at all $\conditionthreshold$ (plasma colormap). 
  }
  \label{fig:cog_combined}
\end{figure}

Figure~\ref{fig:cog_combined} shows the CoG bias against the CHM and the DTM as a function of track count. Against the CHM, the adaptive methods produce sharper profiles than Fourier, since at high $N$, classical Capon and high-$\conditionthreshold$ QSVT converge to a CoG bias of approximately $-\SI{11.5}{\metre}$, while Fourier saturates higher at $\sim\!-\SI{7}{\metre}$. The more negative Capon bias shows sidelobe suppression, since the broad Fourier point-spread function leaks power into high-elevation sidelobes above the canopy. This inflates its power-weighted height, while Capon's adaptive nulling removes this leakage and the centroid settles onto the true two-layer distribution (weighted toward the lower scatterers; ground and ground--trunk double bounce), pulling the power-weighted height downward. The picture against the DTM is complementary since the CoG bias is now measured upward from the ground and the sharper Capon profiles place their centroid \emph{lower} above the terrain. High-$\conditionthreshold$ QSVT and classical converge to $\sim\!+\SI{11}{\metre}$ at high $N$, while the broader Fourier profile spreads power upward and saturates \emph{higher}, at $\sim\!+\SI{15}{\metre}$. The same ordering by profile sharpness therefore appears in both panels, with its sign reversed by the choice of reference. Specifically, Capon pulls the centroid downward, which reads as a more negative bias against the canopy top and a smaller positive bias above the ground.

Another important observation is the behavior of the low-$\conditionthreshold$ curves. Although every $\conditionthreshold$ stays within $\SI{0.4}{\metre}$ of the classical baseline in the peak-height metrics (Sec.~\ref{sec:res_validity}), the CoG metrics fan out for $N\gtrsim 5$. Lowering $\conditionthreshold$ moves the CoG from the classical Capon value toward the Fourier one. At $N=11$ the CoG-versus-CHM bias is about $-\SI{11.5}{\metre}$ for the high-$\conditionthreshold$ cluster, $-\SI{6.7}{\metre}$ at $\conditionthreshold=15$, close to the Fourier value of $-\SI{7.4}{\metre}$, and $-\SI{1.4}{\metre}$ at $\conditionthreshold=8$, which lies beyond the Fourier value for $N\ge 9$. The CoG-versus-DTM bias shows the same ordering, with about $\SI{11}{\metre}$ for the high-$\conditionthreshold$ cluster, $\SI{15.3}{\metre}$ for Fourier, $\SI{15.8}{\metre}$ at $\conditionthreshold=15$ and $\SI{21.4}{\metre}$ at $\conditionthreshold=8$. The mechanism is the selective truncation of Section~\ref{sec:method}.  If $\conditionthreshold$ is too small to cover a pixel's spectrum, the polynomial approximation does not invert the small eigenvalues that Capon relies on for tight sidelobe nulling, but suppresses them instead. Power moves into the sidelobes above the canopy, and as a consequence the CoG rises. Diagonal loading maps the small eigenvalues to a constant and tends towards the Fourier response, whereas the odd polynomial maps them towards zero, so at the smallest bound the CoG moves past the Fourier value instead of settling on it.
 
\begin{figure}[tb]
\centering
\includegraphics[width=1.05\linewidth]{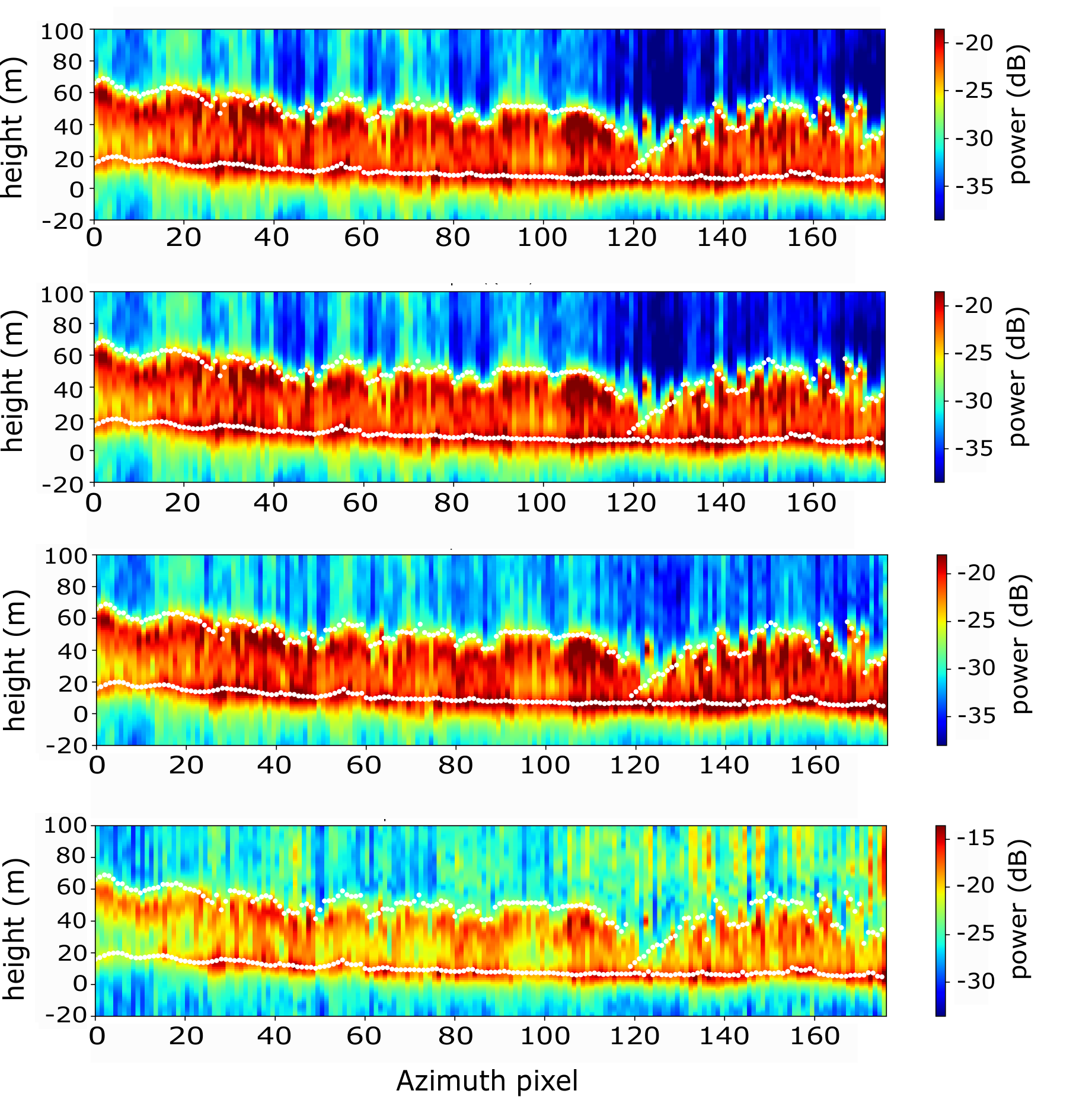}
\caption{%
From top to bottom: classical Capon (reference); quantum Capon at $\conditionthreshold=119$, $37$, and $8$
along the azimuth line of Fig.~\ref{fig:profiles_all} ($N=11$). 
The $\conditionthreshold=8$ panel uses a color scale shifted up by 5~dB. White dots mark the LiDAR DTM and DTM\,$+$\,CHM.
}
\label{fig:kappa_effect}
\end{figure}

Note that this spread appears in the CoG but is \emph{absent} from the peak height. The location of the spectrum maximum, is set by the brightest, well-conditioned eigenvector direction, which is not affected by the polynomial's behavior on small eigenvalues. $\conditionthreshold$-truncation therefore does not change the location of the dominant peak (Fig.~\ref{fig:bias_vs_tracks}) but instead reshapes the profile tails, redistributing power and shifting the power-weighted height. Figure~\ref{fig:kappa_effect} shows this along a single azimuth line, where at fixed $N=11$, the canopy and ground peak locations are visually invariant as $\conditionthreshold$ decreases from the classical reference through $\conditionthreshold=119,\,37,\,8$, while the profile shape degrades progressively. We can see the nulls filling in and the bands losing their sharp edges, so the contrast between peaks and nulls decreases. Also, we can already see that at $\conditionthreshold=119$ the quantum is indistinguishable from classical Capon. By $\bar{\kappa}=8$ the nulls have largely filled in and power extends well above the canopy.

In this dataset, the aggregate peak-height statistics are less sensitive to the bound than the shape of the reconstructed profile. This means that practical applications may not need the full depth $d=\mathcal{O}(\conditionthreshold\log(\conditionthreshold/\varepsilon))$ when only the canopy peak is required. Applications demanding faithful profile reconstruction (e.g. biomass retrieval from the full vertical structure) would instead require high $\conditionthreshold$ and the corresponding deeper circuits. To our knowledge, this is the first time that the condition number bound of a QSVT phase polynomial has been used as an actual regularization knob with a physical meaning that a user can tune, in a real-world signal-processing pipeline. Our proposed mechanism is also different in kind from the diagonal loading of classical adaptive beamforming. It acts selectively on the spectrum rather than shifting it uniformly.

\section{Conclusion}
\label{sec:conclusion}

We have presented a quantum Capon beamformer, which is a quantum realization of the Capon (MVDR) estimator, and validated it on ESA AfriSAR P-band data over the Mondah forest, against the classical Capon and Fourier baselines with LVIS LiDAR as ground truth. The covariance matrix is block encoded, its inverse is implemented through QSVT, and the beamformer power is read out directly from the resulting quantum circuit by a Hadamard test. The full pipeline, from covariance to Capon spectrum, runs as a single quantum circuit. 

Across the sweeps in track count $N$ and condition number bound $\conditionthreshold$, the quantum inversion reproduces the classical Capon tomogram for $\conditionthreshold \ge 148$, with aggregate peak-height RMSE within $\SI{0.02}{\metre}$ of classical Capon and mean SSIM above $0.99$ at every track count.

Our QSVT formulation showed an additional regularization mechanism in addition to diagonal loading. 
The condition number bound of the phase polynomial acts as a spectrally
selective regularizer, where eigenvalues above $1/\conditionthreshold$ are
inverted accurately while smaller ones are suppressed, in contrast to the
uniform shift of diagonal loading. Lowering $\bar{\kappa}$ moves the
reconstruction continuously away from the adaptive Capon response.
Since $\conditionthreshold$ also fixes the interval $[1/\conditionthreshold,1]$
on which the inverse is approximated, it controls the circuit depth and the
spectral regularization simultaneously. Increasing $\conditionthreshold$ widens
this interval and sharpens the reconstruction, at the cost of a longer phase
sequence and a deeper circuit.
At $N=11$ the
CoG reaches the non-adaptive Fourier value at $\bar{\kappa} \approx 15$ and
passes it at $\bar{\kappa}=8$, while the aggregate peak-height RMSE changes by
less than $0.4$\,m. This ``knob'' therefore reshapes the profile tails and
sidelobe structure without relocating the dominant peak. Peak based canopy
height retrieval is therefore robust to aggressive, low-$\conditionthreshold$
truncation, where applications requiring faithful vertical structure require
larger $\conditionthreshold$ and hence deeper circuits. To our knowledge, this
is the first demonstration that the condition number bound of the QSVT phase
polynomial can be used as a user-tunable regularization parameter in a
real-world signal-processing pipeline.

The quantum Capon requires $\bigO{\log N}$ qubits, which is an exponentially smaller working requirements in comparison to the $\bigO{N^2}$ of the classical storage for the covariance matrix and its inverse.
This property makes the quantum Capon approach attractive in principle for the potentially larger covariance matrix dimensions required in related adaptive filtering and super-resolution problems.

There are some potential future directions that could follow from this work. To begin with, one could employ the asymptotically optimal phase construction~\cite{Lin2020} used here to pair it with block encodings tailored for specific hardware~\cite{Camps2024, Snderhauf2024} that exploit the Toeplitz-like structure of multi-baseline covariance matrices, thus reducing the block-encoding cost. 
Additionally, if the acquisition geometry is chosen such that $N=2^n$ and the wavenumbers satisfy Eq.~\eqref{eq:wavenumber-cond}, the steering-state preparation can be reduced to a constant depth.
This suggests that baseline design, although it is currently driven by classical factors, such as ambiguity suppression and coverage, would acquire an additional criterion when the quantum beamformer is employed.

\section*{Acknowledgment}
We would like to thank Sylvia Liberata Ullo and Francesco Mauro from University of Sannio for their assistance with accessing the ESA datasets and classical TomoSAR code and literature necessary for the study. The TomoSAR data preparation and processing pipeline is adapted from the practical materials of M. Pardini and L. Ferro-Famil, \href{https://eo4society.esa.int/resources/7th-advanced-training-course-radar-polarimetry/}{ESA's 7th Advanced Training Course on Radar Polarimetry, 2023}. This activity is part of the “SpinQC4EO” project with number 4000149905/25/I-DT-bgh.

\ifCLASSOPTIONcaptionsoff
  \newpage
\fi

\bibliographystyle{IEEEtran}
\bibliography{bibliography}

\end{document}